\documentclass[aps,pra,twocolumn,groupedaddress]{revtex4-1}

\usepackage{epsfig,amssymb,amsmath,amsfonts}
\usepackage{times}
\usepackage{graphicx}
\usepackage{epstopdf}
\newcommand{\bra}[1]{{\left\langle{#1}\right\vert}}
\newcommand{\ket}[1]{{\left\vert{#1}\right\rangle}}

\def\>{\rangle}
\def\<{\langle}

\newcommand{\ignore}[1]{}

\def\bmat{\left[\begin{matrix}} 
\def\emat{\end{matrix}\right]}

\begin{document}

\title{Comparison of a quantum error correction threshold for exact and approximate errors}

\author{Mauricio Guti\'errez}
\author{Kenneth R. Brown}
\email{ken.brown@chemistry.gatech.edu}
\affiliation{Schools of Chemistry and Biochemistry; Computational Science and Engineering; and Physics\\ 
Georgia Institute of Technology, Atlanta, GA 30332-0400}

\date{\today}

\begin{abstract}

Classical simulations of noisy stabilizer circuits are often used to estimate the threshold of a quantum error-correcting code. Physical noise sources are efficiently approximated by random insertions of Pauli operators. For a single qubit,  more accurate approximations that still allow for efficient simulation can be obtained by including Clifford operators and Pauli operators conditional on measurement in the noise model. We examine the feasibility of employing these expanded error approximations to obtain better threshold estimates.  We calculate the level-1 pseudothreshold for the Steane [[7,1,3]] code for amplitude damping and dephasing along a non-Clifford axis.  The expanded channels estimate the actual channel action more accurately than the Pauli channels before error correction.  However, after error correction, the Pauli twirling approximation yields very accurate estimates of the performance of quantum error-correcting protocols in the presence of the actual noise channel.

\end{abstract}


\maketitle

\section{Introduction}\label{Sec:Intro}

The threshold theorem of quantum error correction promises the accurate implementation of arbitrary size quantum algorithms if the underlying physical errors are below certain values.  The error thresholds depend strongly on the specific quantum error correcting code, how errors are detected and fixed \cite{Poulin_soft_decoding, Goto_soft_decoding, Suchara_leakage}, and what errors are assumed \cite{Novais_correlated_errors, Fowler_leakage, Yu_and_Krysta, Novais_beyond_single_errors}.  Most codes have been designed to fix random Pauli errors and error correction procedures can be simulated efficiently using the stabilizer formalism \cite{Gottesman_thesis, Aaronson}. A broader class of errors including Clifford operations \cite{Cory} and Pauli measurements \cite{PRA_us} can also be included in this formalism. For a single qubit, this extended error set has been shown to yield improved approximations of realistic error models including amplitude damping \cite{PRA_us}. 

Here we examine whether these improved approximations also lead to more accurate threshold estimates. Specifically, we calculate the level-1 pseudothreshold for the Steane [[7,1,3]] code \cite{Steane1996} for two non-stabilizer errors, amplitude damping and a depolarization channel along a magic-state axis,  and compare the exact solution to approximations based on Pauli errors or  Clifford and  Pauli measurement errors. The Steane code has been well studied theoretically \cite{SteanePRA2003,SvoreQIC2007, MetodiMicro2005, TomitaPRA2013,AbuNadaPRA2014,WeinsteinPRA2014,AndersonPRL2014} and a logically encoded state has been recently demonstrated experimentally \cite{NiggScience2014}. The code is small enough to allow for exact simulation similar to recent work on distance-3 surface codes, which compared a realistic error model corresponding to T$_1$ (amplitude damping) and T$_2$ (dephasing) processes and an approximate Pauli error model based on twirling \cite{Yu_and_Krysta}. 

In addition to the pseudothreshold, we are interested in two other qualities of the approximation, the accuracy and the honesty.  The accuracy is a measure of how close is the state generated by the approximate evolution  to the state generated by the exact evolution. We describe an approximation as honest if the final state after the approximate evolution is further from the initial state than the final state after the exact evolution.  In other words, an approximation is honest if it upper-bounds the error of the exact evolution.  As pointed out by Puzzuoli \textit{et al.} the composition of honest approximations is not necessarily honest \cite{Cory2014}.  This implies that an approximation that is honest at the 1-qubit physical level might lead to a dishonest representation of the overall error produced on the system.   If our goal is to employ our approximate channels to infer the performance of error-correcting strategies under realistic non-stabilizer noise, then we need to be cautious and be sure that they compose in an honest fashion.  We provide numerical evidence that, in the context of an error-correcting circuit, an honest approximation at the physical level remains honest at the logical level.  Furthermore, we show that, for the error models studied, physically dishonest approximations based on the Pauli channel might lead to approximations at the logical level that are both approximately honest and very accurate, in agreement with similar results obtained by Geller and Zhou \cite{Geller_and_Zhou}.  This suggests that it might not be necessary for the approximations to be honest at the physical level.  

The paper is organized as follows.  In Section \ref{sec:channels}, we describe the target realistic error channels and our method for generating approximate channels \cite{PRA_us}. In Section \ref{sec:hon_and_acc}, we review the important concepts of honesty and accuracy of an approximate channel.  In Section \ref{sec:threshold_method} we explain our procedure for calculating the pseudothreshold.  In Section \ref{sec:results}, we present our results before concluding in Section \ref{sec:conclusions}. 

\section{Error channels}\label{sec:channels}

We review all the error channels introduced in \cite{PRA_us}.  We start with the stabilizer expansions to the Pauli channel (PC) that are used as models to approximate realistic non-stabilizer error channels.  Next we discuss two important cases of these error channels that lie outside the stabilizer formalism.  Finally we review two different constraints under which the approximations are performed.  All the error channels introduced in this section will be expressed in the operator-sum representation.  

Throughout the paper we use $X$, $Y$, and $Z$ to represent the Pauli matrices with associated eigenvectors $\{|+\>, |-\>\}$, $ \{|+i\>, |-i\>\}$, and $ \{|0\>, |1\>\}$ respectively.

\subsection{Efficiently simulable processes}

One of the main ideas introduced in \cite{PRA_us} was that the PC can be greatly expanded without becoming non-stabilizer. We do this by adding Kraus operators that correspond to either Clifford operations or Pauli measurements followed by conditional Pauli operations.  For the 1-qubit case, the Clifford channel (CC), which corresponds to the first expansion, is composed of the 24 operators that maintain the symmetry of the chiral Clifford octahedron \cite{vanDam2009}. 
Another expansion to the PC can be obtained by introducing pairs of operators that effectively produce a measurement in a Pauli basis followed by a conditional Pauli operation, such that all states are mapped to the same state.  We refer to these pairs of operators as measurement-induced translations.  For each Pauli state, $\ket{\lambda}$, these operations can be represented by the following pairs:
\begin{equation}
\Bigl \lbrace E_{\lambda0} = \ket{\lambda} \bra{\lambda} , E_{\lambda1} = \ket{\lambda}\bra{\lambda^\perp} \Bigl \rbrace 
\end{equation}

The addition of these measurement-induced translations to the PC gives rise to the Pauli + measurements channel (PMC), while their addition to the CC produces the Clifford + measurements channel (CMC).

\subsection{Non-stabilizer error channels}

As in \cite{PRA_us}, we use the efficiently simulable channels to approximate realistic non-stabilizer channels.  In particular, we focus on the amplitude damping channel (ADC) and the polarization along an axis in the x-y plane of the Bloch sphere (Pol$_{\phi}$C), shown on equations \ref{eq:adc} and \ref{eq:polc}, respectively: 

\begin{equation} 
\text{ADC} =
\begin{cases}
E_{A0} = |0\> \<0| \medspace + \medspace \sqrt{1-\gamma} \thinspace |1\> \<1| \\
E_{A1} = \sqrt{\gamma} \thinspace|0\> \<1|
\end{cases} 
\label{eq:adc}
\end{equation}

\begin{equation}
\text{Pol}_{\phi}\text{C} = 
\begin{cases}
E_{xy0} = \sqrt{1-p_\phi} \thinspace I \\
E_{xy1} = \sqrt{p_\phi} \thinspace [\cos(\phi) \thinspace X \medspace  + \medspace  \sin(\phi) \thinspace  Y]
\end{cases}
\label{eq:polc}
\end{equation} 

\subsection{Constraints}

The free parameters in the PC and its expansions correspond to the probabilities associated with the Kraus operators.  Previously, we have obtained stabilizer approximations to the realistic non-stabilizer channels by minimizing the Hilbert-Schmidt distance \cite{Distance_meas, Chuang_distance} over the parameter space of the models.  As we do not want to underestimate the deleterious effect of the target channel on quantum information (i.e. we want our approximation to be ``honest'' \cite{Cory}), we perform the minimizations such that the approximate channels constitute an upper bound on the error induced on the system.  Mathematically, this condition can be enforced in a variety of ways, by employing different fidelity or distance measures.  From the constraints that we have studied, the most lenient one corresponds to the average fidelity constraint, in which we enforce the following condition:
\begin{equation}
F_{\textrm{av}}(I,\text{Target}) \geqslant F_{\textrm{av}}(I,\text{Model})
\end{equation} 
The average fidelity between a unitary transformation $V$ and a quantum channel $K$ is given by:
\begin{equation}
F_{\textrm{av}}(V, K) = \frac{1}{N^2} \sum_{i} | \text{Tr}(V^\dagger K_{i}) |^2
\end{equation}
where $N$ is the dimension of the Hilbert space and $\lbrace K_{i} \rbrace$ are these Kraus operators of the error channel $K$.  On all the approximations that we have performed, the average fidelity constraint has always given the same results as if no constraint had been applied.

On the other hand, the most stringent constraint corresponds to the worst trace distance one, in which we enforce that for every initial pure state its trace distance to the resulting state after the target transformation is not greater than its trace distance to the resulting state after the model approximation:
\begin{equation}
D^{\textrm{Tr}}(\rho, \text{Target}(\rho)) \leq D^{\textrm{Tr}}(\rho, \text{Model}(\rho))
\end{equation}
where the trace distance is calculated using the following expression:
\begin{equation}
D^{\textrm{Tr}}(\rho, \sigma) = \frac{1}{2} \textrm{Tr} |\rho-\sigma|
\end{equation}

This worst trace distance constraint results in approximate channels that are honest, in the sense that the deleterious effect of the target error channel on any pure quantum state will never be underestimated, as pointed out by Magesan $\textit{et al.}$ \cite{Cory} and Puzzuoli $\textit{et al.}$ \cite{Cory2014}. We could think of an even tighter constraint in which we enforce this condition on every initial state, pure or mixed. However, if the target and the model transformations have different fixed points, this condition is impossible to satisfy.  We compare results for both constraints and label them ``a'' (average fidelity constraint) and ``w'' (worst trace distance constraint).

\section{Honesty and accuracy at the physical and logical levels}\label{sec:hon_and_acc}
\begin{figure*}
\centering
\includegraphics[scale=1]{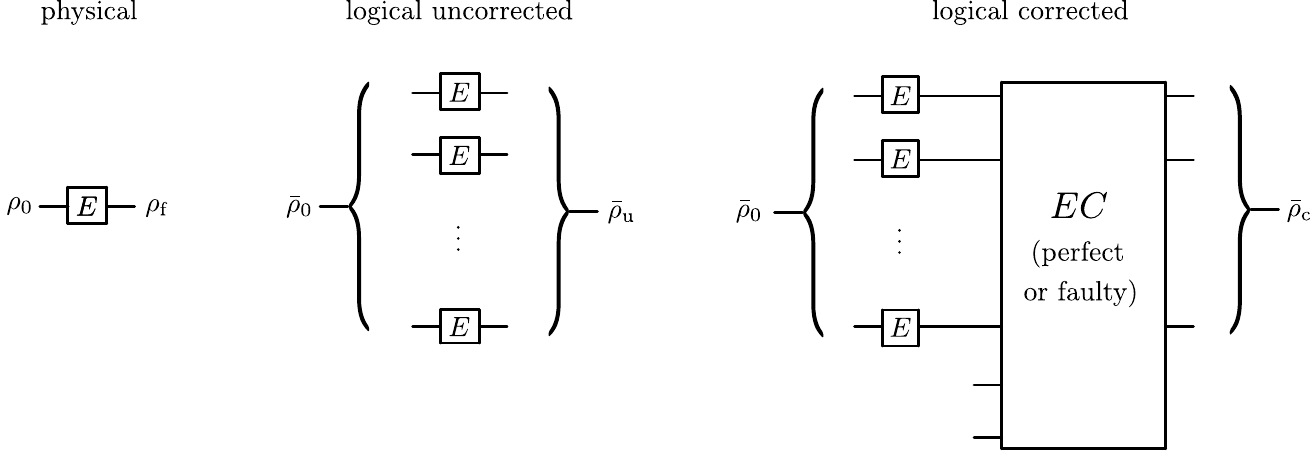}
\caption{We study the honesty and accuracy of the approximate channels at the physical level and at the logical level before and after error correction.  The logical initial state is encoded without errors and then errors are applied.  The preparation of the ancillary cat state in the faulty EC is error-free as well, as shown in Fig. \ref{fig:EC_circuits}.}
\label{fig:the_levels}
\end{figure*}

For each approximate channel, we study two properties: honesty and accuracy. An approximate channel is honest if it does not underestimate the detrimental effect of the target error channel.  The accuracy of an approximate channel refers to how closely it can mimic the effect of the target channel on an initial state.  More explicitly, if a target error channel $E$ maps a pure state $\rho$ to $E(\rho)$ and an approximate channel $A$ maps the same state to $A(\rho)$, then $A$ is honest if 
\begin{equation}
D^{\textrm{Tr}}(\rho , E(\rho)) \leq D^{\textrm{Tr}}(\rho, A(\rho))
\end{equation} 
for every pure state in the initial physical or logical space.  The accuracy is measured by the average trace distance between the resulting states: 
\begin{equation}
\Bigl \langle D^{\textrm{Tr}}(E(\rho) , A(\rho)) \Bigl \rangle
\end{equation}

Notice that for both properties, our measure of choice is the trace distance.  A good approximate channel will be honest (or as least dishonest as possible) and as accurate as possible, not only at the physical level, but also at the various logical levels. We distinguish 4 different scenarios to compare honesty and accuracy: (a) the physical (1-qubit) level, (b) the uncorrected logical level, (c) the logical level with perfect EC, and (d) the logical level with faulty EC, as depicted in Fig. \ref{fig:the_levels}.

For each target non-Clifford error channel, we study two kinds of approximations: (a) the Pauli channels (PC), which employ only 1-qubit Pauli operators, and (b) the expanded channels or Clifford+measurements channels (CMC), which include all the 1-qubit Clifford operators and the measurement-induced translations \cite{PRA_us}.  In turn, each kind of approximation is performed with the average fidelity constraint (``a'') and the worst trace distance constraint (``w''), resulting in four approximate channels.  Notice that the unconstrained PC is equivalent to the Pauli Twirled Approximation \cite{Cory, Cory2014, Geller_and_Zhou}, the channel obtained by removing the off-diagonal elements from the target channel's process matrix in the Pauli basis \cite{Pauli_twirling_Chuang}.  We also analyze the completely isotropic Pauli channel or depolarizing channel (DC), the most common error model used when calculating thresholds. In this paper we are comparing single qubit error channels and we only use the single qubit depolarizing channel. This channel is a version of the PC a where the coefficients corresponding to each Pauli matrix are forced to have the same value.  This error model serves as a reference.  The approximations are summarized in Table \ref{table:channels_summary}.
\begin{table}[htdp]
\caption{Summary of the various target and approximate channels.}
\begin{center}
\begin{tabular}{| c || c | c|}
\hline
\multicolumn{1}{| c ||}{Channel} & \multicolumn{1}{| c |}{Complete name} & \multicolumn{1}{| c |}{Honesty constrained} \\ \cline{1-3}
ADC & amplitude damping & -- \\ \hline
PolC & polarization along non-Clifford axis & -- \\ \hline
PC a & Pauli & no  \\ \hline
PC w & Pauli & yes  \\ \hline
CMC a & Clifford+measurements & no  \\ \hline
CMC w & Clifford+measurements & yes \\ \hline
DC & Depolarizing channel & no \\ \hline
\end{tabular}
\end{center} \label{table:channels_summary}
\end{table}

\section{Calculation of the pseudo-threshold}\label{sec:threshold_method}
Our objective with respect to the pseudo-threshold is twofold.  On the one hand, we want to study how sensitive a QECC's threshold is to the noise model.  On the other hand, we want to determine if the thresholds obtained with our expanded error models approximate the realistic threshold more accurately than the PC.  

\subsection{Procedure to compute the level-1 pseudo-threshold}
Because our target error models are non-stabilizer, we perform exact (full density matrix) simulations of QEC circuits up to the first level of encoding.  We calculate a particular QECC's level-1 pseudo-threshold under a given error channel in the following way:

\begin{enumerate}
  \item Run the physical circuit:
  \begin{enumerate}
    \item Choose an initial 1-qubit pure state, $|\psi \rangle$.
    \item Apply the selected error channel.
    \item Compute the fidelity between the initial and final states.
  \end{enumerate} 

  \item Run the logical circuit: 
  \begin{enumerate}
    \item Encode the initial state using the selected QECC.
    \item Apply the error channel to each physical qubit.
    \item Perform EC.
    \item Compute the fidelity between the initial and final logical states. 
  \end{enumerate}
  
  We are interested in how much the final state is affected by errors which are uncorrectable by the selected QECC.  Therefore, for the faultily corrected case, before computing the fidelity, we perform one round of perfect EC. This has the effect of eliminating correctable errors which happened during or after the faulty EC.  The process of performing a round of perfect EC and then computing the fidelity can also be viewed as computing an error-corrected fidelity:  
  \begin{equation}
  F_{\textrm{EC}}(| \psi_{L} \rangle, \rho_{L}) = \sqrt{\sum_{i} \langle \psi_{L} | E_{i}^{\dagger} P_{i}^{\dagger} \rho_{L} P_{i} E_{i} | \psi_{L} \rangle}  
  \end{equation}
  where $| \psi_{L} \rangle$ is the initial logical state and $\rho_{L}$ is the final logical state, which may not be pure.  The set $\lbrace E_{i} \rbrace$ consists of all error operators which the QECC is designed to correct, while $\lbrace P_{i} \rbrace$ is the set of projectors to the subspaces associated with each error.  For the Steane [[7,1,3]] code, the set of correctable errors is formed by the 64 Pauli operators formed by all possible combinations of $X$ and $Z$ errors acting independently on at most one qubit, which includes the Identity operator for the case of no errors.   

 The set $\lbrace E_{i} \rbrace$ includes all (error) operators which the QECC is designed to correct.  When acting on a given state, $| \psi_{L} \rangle$, in the codespace, each one of these operators will transform it into the equivalent state on each logical subspace.  For example, for the Steane [[7,1,3]] code, the set of correctable errors are the  64 Pauli operators formed by all possible combinations of $X$ and $Z$ errors acting independently on at most one qubit, which includes the Identity operator for the case of no errors.   
  
  \item Repeat steps (1) and (2) for various noise strengths to obtain fidelities for the physical and logical circuits. The threshold is given by the first intersection between the two curves.
  \item Repeat this procedure for several initial states to obtain an average threshold.  For the perfectly corrected case, we select 80 initial points uniformly distributed on the Bloch sphere.  For the faultily corrected one, we select 20.
\end{enumerate}

\subsection{Methods of error correction}

\begin{figure*}[]
\centering
\includegraphics[scale=0.80]{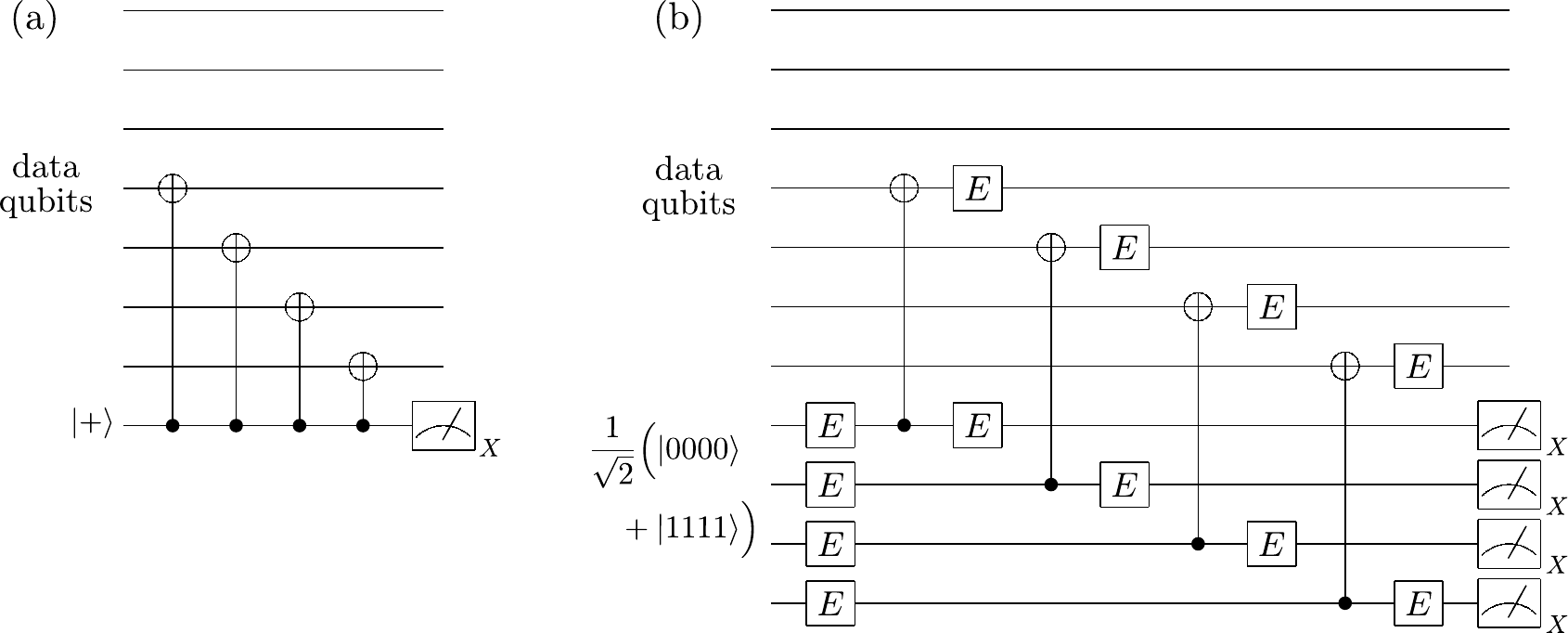}
\caption{Circuits representing the measurement of the operator $IIIXXXX$ in an (a) error-free regime and a (b) faulty regime.  In the former case, we only need to employ 1 ancillary qubit. Notice that the ancillary qubit starts in the $\ket{+}$ state and the measurement is performed in the $X$ basis.  In the faulty EC regime, in order to make the procedure fault tolerant, we employ 4 ancillary qubits initialized in a cat state \cite{DivincenzoPRL1996}.  We then measure each ancillary qubit in the $X$ basis and compute their parity to extract the outcome.}
\label{fig:EC_circuits}
\end{figure*}

The EC step is performed by measuring the stabilizer generators and later correcting any detected errors.  We distinguish between the error-free EC, which results in a code-capacity pseudo-threshold, and the faulty EC, which results in the more realistic circuit-based pseudo-threshold.  The faulty EC is built by inserting an error channel after each gate in the original circuit.  As the Steane code will be the focus of our analysis, consider, for example, the measurement of the stabilizer $IIIXXXX$, as depicted in Fig. \ref{fig:EC_circuits}.  The error-free EC step would consist of circuits analogous to (a) for each stabilizer generator.  On the other hand, in the faulty EC regime, each stabilizer generator would be measured as shown in (b).  Each stabilizer measurement is then repeated and the syndrome is compared to the one in the previous round. If there is a disagreement between these two, a third round of stabilizer measurements is performed and its syndrome is selected as the definitive one.

\section{Results}\label{sec:results}

\subsection{Honesty and accuracy of the approximations}
By construction, the ``w'' approximations are honest at the physical (1-qubit) level, provided that the initial state is pure.  In our previous work we also determined that when approximating a general non-Clifford channel at the physical level, the expanded channels are more accurate than the Pauli.  Before computing the level-1 pseudo-thresholds for different approximations, we first examine if the honesty of the ``w'' approximations and the greater accuracy of the expanded channels were maintained at the logical level.

\subsubsection{Amplitude Damping Channel (ADC)}

\begin{figure*}[h!]
\centering
\includegraphics[scale=0.35]{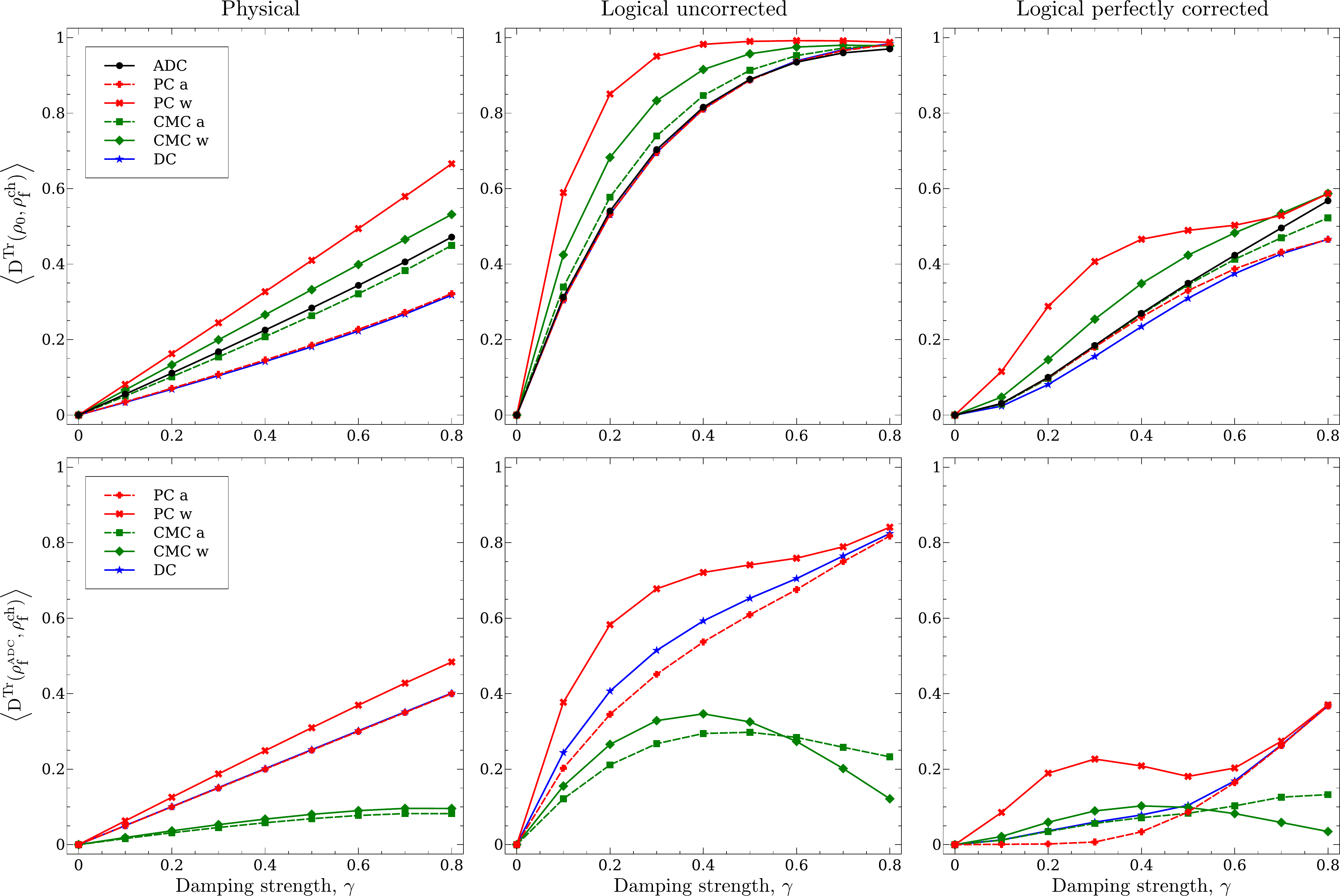}
\caption{(color online) Honesty (top) and accuracy (bottom) of the approximate channels to the ADC at various levels.}
\label{fig:ADC_honesty}
\end{figure*}

For the physical, logical uncorrected, and logical with perfect EC levels, we have selected 80 initial states uniformly distributed over the Bloch sphere surface.  For the logical faultily corrected level, we have selected 20 points, as the simulations involve 3 extra qubits and consequently take an exponentially longer time.  We have computed the trace distance between each one of them and the resulting final state after the ADC and its approximations.  The average distances are shown in the first row of Fig. \ref{fig:ADC_honesty} as a function of the damping strength, $\gamma$.  Likewise, we have computed the trace distance between each final state after the ADC and each final state after every approximate channel.  The average distances are presented in the second row of Fig. \ref{fig:ADC_honesty}. The behavior in the limit of small damping strength ($\gamma \rightarrow 0$) is summarized in Tables \ref{table:Honesty_ADC_small_gamma} and \ref{table:Accuracy_ADC_small_gamma}.  In this limit, it is useful to Taylor-expand the distances in terms of the noise strength and compare the coefficients of the leading order terms. Expectedly, for the corrected logical cases the linear term is suppressed and the leading order is quadratic.  For the physical and uncorrected logical cases, the leading order is linear.  At the logical level with faulty EC, simulations were only carried out at low damping strengths $\bigl ( \gamma \in [10^{-5}, 10^{-3}] \bigl )$, which is the pertinent region for the pseudo-threshold computation.            

\begin{table*}[htdp]
\caption{Honesty of the approximate channels to the ADC in the limit of small damping strength. Standard deviations below 10$^{-9}$ are not reported.}
\begin{center}
\begin{tabular}{| c || c | c | c | c |}
\hline
\multicolumn{1}{| c ||}{Channel} & \multicolumn{1}{| c |}{Physical} & \multicolumn{1}{| c |}{Logical uncorrected} & \multicolumn{1}{| c |}{Logical perfectly corrected} &
\multicolumn{1}{| c |}{Logical faultily corrected} \\ \cline{2-5}
& D$^{\textrm{Tr}} / \gamma$ & D$^{\textrm{Tr}} / \gamma$ 
& D$^{\textrm{Tr}} / \gamma^2$ & D$^{\textrm{Tr}} / (10^{2} \gamma^{2})$ \\ \hline
ADC & $0.55(27)$ & $3.62$ & $3.76(96)$ & $8.0(1.8)$ \\ \hline
PC a & $0.347(79)$ & $3.50$ & $3.76(96)$  & $7.8(1.8)$   \\ \hline
PC w & $0.81(12)$   & $8.35$  & $18.5(3.5)$  & $37.7(8.0)$  \\ \hline
CMC a & $0.50(18)$  & $4.00$  & $3.48(45)$  & $6.3(1.2)$ \\ \hline
CMC w & $0.66(24)$  & $5.33$  & $6.19(80)$  & $11.3(2.2)$ \\ \hline
DC & $0.333$ & $3.50$  & $2.75(36)$  & $4.95(96)$ \\ \hline
\end{tabular}
\end{center} \label{table:Honesty_ADC_small_gamma}
\end{table*}

\begin{table*}[htdp]
\caption{Accuracy of the approximate channels to the ADC in the limit of small damping strength.  Standard deviations below 10$^{-9}$ are not reported.}
\begin{center}
\begin{tabular}{| c || c | c | c | c |}
\hline
\multicolumn{1}{| c ||}{Channel} & \multicolumn{1}{| c |}{Physical} & \multicolumn{1}{| c |}{Logical uncorrected} & \multicolumn{1}{| c |}{Logical perfectly corrected} &
\multicolumn{1}{| c |}{Logical faultily corrected} \\ \cline{2-5}
& D$^{\textrm{Tr}} / \gamma$ & D$^{\textrm{Tr}} / \gamma$ 
& D$^{\textrm{Tr}} / \gamma^2$ & D$^{\textrm{Tr}} / (10^{2} \gamma^2)$ \\ \hline
PC a & $0.500$ & $2.41$ & $7(12) \times 10^{-6}$ & $0.123(28)$ \\ \hline
PC w & $0.63(26)$ & $4.94$ & $14.8(2.6)$ & $29.8(6.2)$ \\ \hline
CMC a & $0.166(60)$ & $1.35$ & $1.61(44)$ & $2.15(74)$ \\ \hline
CMC w & $0.194(60)$ & $1.75$ & $3.05(94)$ & $3.7(1.1)$ \\ \hline
DC & $0.505(97)$ & $2.92$ & $1.68(69)$ & $3.2(1.2)$ \\ \hline
\end{tabular}
\end{center} \label{table:Accuracy_ADC_small_gamma}
\end{table*}

Notice that at the physical level in the first row of Fig. \ref{fig:ADC_honesty}, the ``w'' approximations result in curves that are above the target ADC by construction, while the ``a'' approximations produce curves below it.  This behavior is also present in the small noise strength limit, as can be seen by the magnitudes of the linear coefficients (Table \ref{table:Honesty_ADC_small_gamma}): PC a $<$ CMC a $<$ ADC $<$ CMC w $<$ PC w.   Likewise, the accuracies of the CMC approximations are much better than that of the PC approximations (Table \ref{table:Accuracy_ADC_small_gamma}). In the $\gamma \rightarrow 0$ limit, the CMC approximations are $\approx 3$ times more accurate.        

At the three logical levels, the ``w'' approximations are honest for every damping strength. This is true not just on average, but for every initial state considered.  This is an important result, as it means that we can safely use the ``w'' approximations as a substitute of the ADC when determining codes' thresholds or other error-correcting properties.  Remarkably, the dishonesty of the PC a is greatly reduced from the physical to the logical levels  in the limit of small $\gamma$.  Its error is below the honesty cutoff by $36\%$ at the physical level but by less than only 2\%  for both corrected logical levels and well within the deviation in the distance.  In contrast, the dishonesty of the CMC a is not improved at the logical levels and is below the honesty cutoff by $8-20\%$ for all cases. 

The variation of the accuracy from level to level shows an interesting behavior. For both levels where the effect of the errors is linear (physical and uncorrected logical), in general the CMC channels and the ``a'' approximations are more accurate than the PC channels and the ``w'' approximations, respectively.  This is seen by the magnitudes of the linear coefficients (Table \ref{table:Accuracy_ADC_small_gamma}): CMC a $<$ CMC w $<$ PC a $<$ PC w.  At the logical level with perfect EC, this intuitively expected behavior is seen only for high damping strengths ($\gamma > 0.5$) (see Fig. \ref{fig:ADC_honesty}).  Surprisingly, for lower damping strengths, the most accurate approximation is given by the unconstrained PC, as can be observed by the suppresion of the second order terms in the accuracy ( Table \ref{table:Accuracy_ADC_small_gamma}).  This behavior is particularly pronounced at the logical level with perfect EC, where the second order terms for the PC a and ADC are practically indistinguishable.

\subsubsection{Polarization along a non-Clifford Axis Channel (Pol$_{\phi}$C)}

\begin{figure*}[h!]
\centering
\includegraphics[scale=0.35]{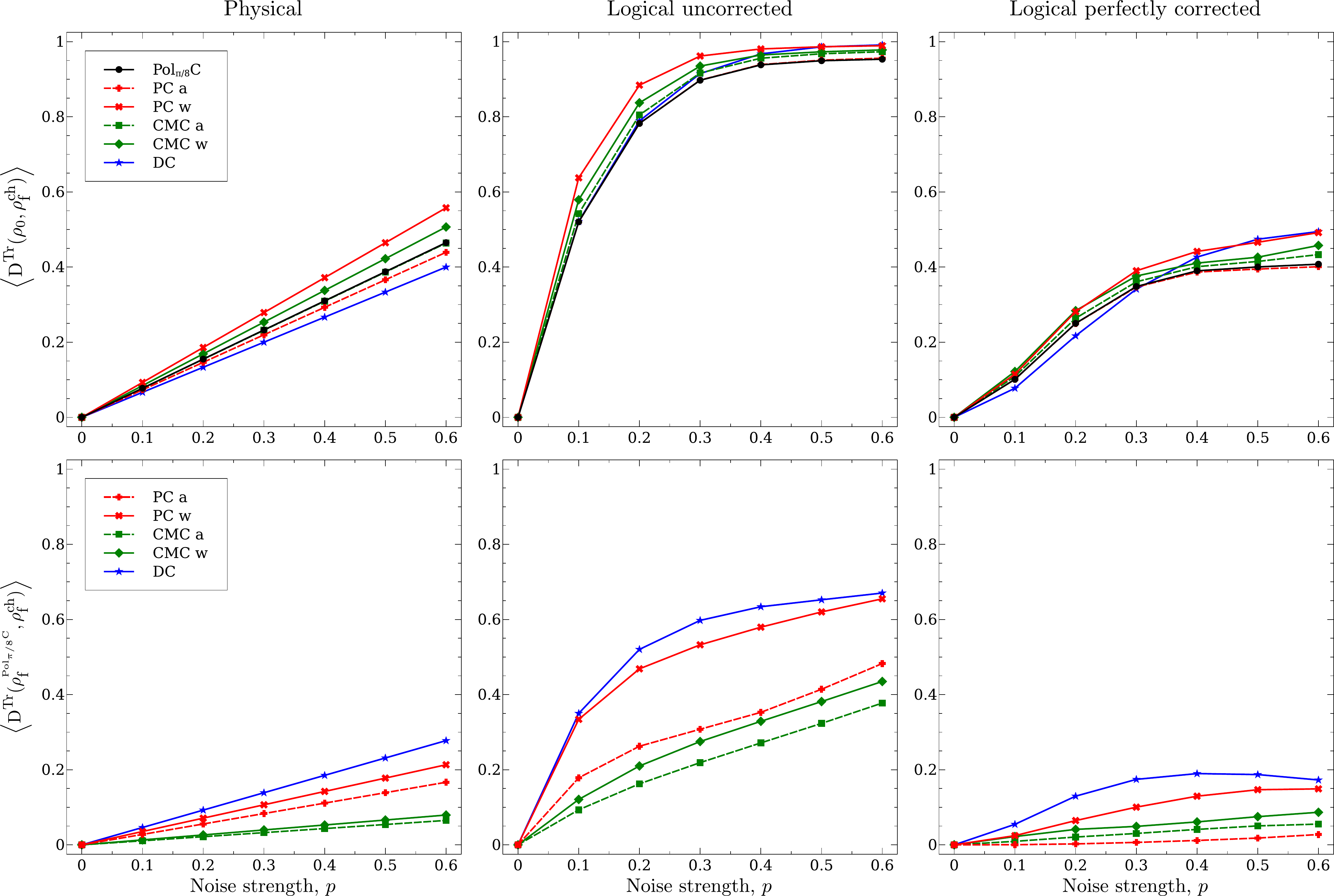}
\caption{(color online) Honesty (top) and accuracy (bottom) of the approximate channels to the Pol$_{\pi/8}$C at various levels.}
\label{fig:PolXY_honesty}
\end{figure*}

We perform an analogous analysis for our second target error channel: the polarization along a non-Clifford axis on the $XY$ plane of the Bloch sphere.  We select the axis forming an angle $\phi = \pi/8$ with respect to the $X$ axis, as this is the angle for which the expanded error models perform the worst \cite{PRA_us}.  Once again, we have selected 20 initial states for the faultily corrected level and 80 points for all other levels.  We have computed the trace distance between each one of them and the resulting final state after the Pol$_{\pi/8}$C and its approximations.  The average distances are shown in the first row of Fig. \ref{fig:PolXY_honesty} as a function of the noise strength, $p$.  Likewise, we have computed the trace distance between each final state after the Pol$_{\pi/8}$C and each final state after every approximate channel.  The average distances are presented in the second row of Fig. \ref{fig:PolXY_honesty}. The behavior in the limit of small noise strength ($p \rightarrow 0$) is summarized in Tables \ref{table:Honesty_PolXY_small_p} and \ref{table:Accuracy_PolXY_small_p}. As for the ADC, at the physical and uncorrected logical levels, the leading order is linear.  At the corrected logical levels, the leading order is quadratic.           

\begin{table*}[htdp]
\caption{Honesty of the approximate channels to the Pol$_{\pi/8}$C in the limit of small noise strength. Standard deviations below 10$^{-9}$ are not reported.}
\begin{center}
\begin{tabular}{| c || c | c | c | c |}
\hline
\multicolumn{1}{| c ||}{Channel} & \multicolumn{1}{| c |}{Physical} & \multicolumn{1}{| c |}{Logical uncorrected} & \multicolumn{1}{| c |}{Logical perfectly corrected} &
\multicolumn{1}{| c |}{Logical faultily corrected} \\ \cline{2-5}
& D$^{\textrm{Tr}} / p$ & D$^{\textrm{Tr}} / p$ 
& D$^{\textrm{Tr}} / p^2$ & D$^{\textrm{Tr}} / (10^{3} p^2)$ \\ \hline
Pol$_{\pi/8}$C & $0.78(24)$ & $7.00$ & $16.2(4.7)$ & $2.22(72)$ \\ \hline
PC a & $0.73(20)$ & $7.00$ & $16.2(4.7)$ & $2.22(72)$ \\ \hline
PC w & $0.93(18)$ & $9.47$ & $18.5(4.9)$ & $2.71(77)$ \\ \hline
CMC a & $0.77(22)$ & $7.41$ & $17.8(5.1)$ & $2.67(83)$ \\ \hline
CMC w & $0.84(23)$ & $8.17$ & $20.8(5.9)$ & $3.12(96)$ \\ \hline
DC & $0.667$ & $7.00$ & $11.0(1.4)$ & $1.82(35)$ \\ \hline
\end{tabular}
\end{center} \label{table:Honesty_PolXY_small_p}
\end{table*}

\begin{table*}[htdp]
\caption{Accuracy of the approximate channels to the Pol$_{\pi/8}$C in the limit of small noise strength. Standard deviations below 10$^{-9}$ are not reported.}
\begin{center}
\begin{tabular}{| c || c | c | c | c |}
\hline
\multicolumn{1}{| c ||}{Channel} & \multicolumn{1}{| c |}{Physical} & \multicolumn{1}{| c |}{Logical uncorrected} & \multicolumn{1}{| c |}{Logical perfectly corrected} &
\multicolumn{1}{| c |}{Logical faultily corrected} \\ \cline{2-5}
& D$^{\textrm{Tr}} / p$ & D$^{\textrm{Tr}} / p$ 
& D$^{\textrm{Tr}} / p^2$ & D$^{\textrm{Tr}} / (10^{2} p^2)$ \\ \hline
PC a & $0.278(82)$ & $2.47$ & $6.1(7.7) \times 10^{-6}$ & $5.8(1.6) \times 10^{-5}$ \\ \hline
PC w & $0.36(21)$ & $4.95$ & $3.8(1.1)$ & $6.2(1.6)$ \\ \hline
CMC a & $0.108(25)$ & $1.09$ & $1.76(41)$ & $4.7(1.0)$ \\ \hline
CMC w & $0.132(43)$ & $1.37$ & $4.6(1.1)$ & $9.2(2.4)$ \\ \hline
DC & $0.46(11)$ & $4.67$ & $9.2(1.3)$ & $9.7(1.5)$ \\ \hline
\end{tabular}
\end{center} \label{table:Accuracy_PolXY_small_p}
\end{table*}

As observed on the ADC, the ``w'' approximations are honest at every level and for every noise strength. This holds in the average case and also for each initial state considered.  Interestingly, the CMC a becomes honest on average and the PC a average distances are indistinguishable from honest.

Notice that just like for the ADC, at the physical and uncorrected logical levels, the CMC channels and the ``a'' approximations are more accurate than their counterparts PC and ``w'', respectively, as can be seen by the magnitudes of the linear coefficients (Table \ref{table:Accuracy_PolXY_small_p}):  CMC a $<$ CMC w $<$ PC a $<$ PC w.  At the physical level, and in the $p \rightarrow 0$ limit, the CMC approximations are $\approx 3$ times more accurate than the PC approximations. At the corrected logical levels, the most accurate approximation is once again given by the PC a. Surprisingly, this behavior holds even up to high noise strengths ($p = 0.6$).  In the low noise limit, and at the corrected logical levels, the second order terms are practically suppressed.  In this limit, the PC a is on average more accurate than the CMC channels  by a factor of 10$^5$.

\subsubsection{High accuracy of the unconstrained PC}

For both the ADC and the Pol$_{\pi/8}$C, the unconstrained PC results in approximations that are honest (or almost honest) and extremely accurate at the logical corrected levels. In the limit of small error, this is very evident by comparing the quadratic coefficients of the PC a to the other approximations (See Tables \ref{table:Accuracy_ADC_small_gamma}  and \ref{table:Accuracy_PolXY_small_p}.).  The high accuracy of the unconstrained PC in the context of EC has previously been observed. Geller and Zhou found very good agreement between the PC a or Pauli twirled approximation and two different realistic noise models when correcting a Bell state \cite{Geller_and_Zhou}.  Likewise, Puzzuoli \textit{et al.} observed great accuracy of the PC a when correcting a Choi state encoded in the [[5,1,3]] code \cite{Cory2014}.  As clearly explained in this Ref. \cite{Cory2014}, after (perfect) EC, the process matrix elements corresponding to Pauli error strings that result in different syndromes become zero.  Intuitively, we can say that the ``non-Pauli'' advantage of the expanded approximations at the physical level gets ``washed away'' after EC.

\subsection{Level-1 pseudo-thresholds}

We perform the simulations of two different scenarios: (1) one with perfect EC, which results in a relatively high code-capacity pseudo-threshold and (2) one with faulty EC, which results in a more realistic circuit-based pseudo-threshold.  Apart from the average pseudo-threshold, we also calculate the root mean square difference (RMS) between the pseudo-thresholds given by the target non-Clifford channel and the ones predicted by each approximate channel:
\begin{equation}
\textrm{RMS} = \sqrt{\langle (\textrm{p}_{th}^{\textrm{channel}} - \textrm{p}_{th}^{\textrm{approx}})^{2} \rangle}
\end{equation}   

The RMS quantifies the accuracy of each approximate channel to estimate the pseudo-threshold of the target channel.  We calculate the RMS because comparing only the average values does not account for any cancellation of errors. A certain approximate channel can do a very poor job at approximating the pseudo-threshold for every initial state, but result in an average that is close to the target's average.    

\begin{table}[htdp]
\caption{Thresholds for the Steane code under the ADC and its Pauli and expanded approximations. ADC/PC a uses  ADC at the physical level and PC a at the logical level. }
\begin{center}
\begin{tabular}{| c || c | c | c | c |}
\hline
\multicolumn{1}{| c ||}{Channel} & \multicolumn{2}{| c |}{Code capacity} & \multicolumn{2}{| c |}{Circuit-based} \\ \cline{2-5}
 & $\langle \gamma_{th} \rangle$ & RMS & $\langle \gamma_{th} \rangle \times 10^{4}$ & RMS $\times 10^{4}$ \\ \hline
ADC & $0.18(17)$ & -- & $4.8(4.2)$  & -- \\ \hline
PC a & $0.132(38)$ & $0.171$ & $4.8(1.4)$ & $3.91$  \\ \hline
PC w & $0.061(43)$ & $0.204$ & $2.36(60)$ & $4.69$ \\ \hline
CMC a & $0.19(17)$ & $0.0498$ & $6.4(4.2)$ & $1.67$  \\ \hline
CMC w & $0.15(14)$ & $0.0644$ & $4.8(3.1)$ & $1.12$ \\ \hline
DC & $0.162(22)$ & $0.165$ & $7.2(1.4)$ & $4.60$ \\ \hline
ADC/PC a & $0.30(37)$ & $0.255$ & $4.9(4.2)$ & $0.101$  \\ \hline
\end{tabular}
\end{center} \label{table:summarySteaneADC}
\end{table}

\begin{table}[htdp]
\caption{Thresholds for the Steane code under the Pol$_{\pi/8}$C and its Pauli and expanded approximations.  Pol$_{\pi/8}$C/PC a uses  Pol$_{\pi/8}$C at the physical level and PC a at the logical level.}
\begin{center}
\begin{tabular}{| c || c | c | c | c | c |}
\hline
\multicolumn{1}{| c ||}{Channel} & \multicolumn{2}{| c |}{Code capacity} & \multicolumn{2}{| c |}{Circuit-based} \\ \cline{2-5}
 & $\langle p_{th} \rangle$ & RMS & $\langle p_{th} \rangle \times 10^{4}$ & RMS $\times 10^{4}$ \\ \hline
Pol$_{\pi/8}$C & $0.14(24)$ & -- & $3.5(1.5)$  & -- \\ \hline
PC a & $0.086(74)$ & $0.238$ & $3.10(26)$ & $1.53$ \\ \hline
PC w & $0.078(16)$ & $0.237$ & $3.46(35)$ & $1.48$ \\ \hline
CMC a & $0.11(21)$ & $0.112$ & $3.09(85)$ & $0.816$ \\ \hline
CMC w & $0.09(14)$ & $0.169$ & $2.91(76)$  & $0.991$ \\ \hline
DC & $0.083(12)$ & $0.240$ & $3.92(64)$ & $1.60$ \\ \hline
Pol$_{\pi/8}$C/PC a & $0.14(25)$ & $0.0255$ & $3.5(1.5)$ & $1.19 \times 10^{-3}$ \\ \hline
\end{tabular}
\end{center} \label{table:summarySteanePolXY}
\end{table}

The results for the ADC and the Pol$_{\pi/8}$C are summarized in Tables \ref{table:summarySteaneADC} and \ref{table:summarySteanePolXY}, respectively.  For the ADC, the pseudo-thresholds are expressed in terms of the damping strength, $\gamma$, while for the Pol$_{\pi/8}$C, they are expressed in terms of the noise strength, $p$.  In both cases, the standard deviation of the pseudo-thresholds is included inside parentheses.  Notice that the code-capacity pseudo-thresholds are about 3 orders of magnitude higher than the circuit-based ones. The latter ones are on the range expected for the Steane code \cite{Steane1997}.  Although the code-capacity pseudo-thresholds are unrealistically high, they show similar trends with respect to their circuit-based counterparts.

In general, the target non-Clifford channels and their CMC approximations result in high standard deviations of the pseudo-thresholds around its average values.  This implies that for these channels the pseudo-thresholds are much more dependent on the initial state than for the PCs.  It is interesting that the Pauli channels always result in pseudo-thresholds that are lower than the real ones.  This trend has also been observed by Tomita and Svore on the surface code \cite{Yu_and_Krysta} and suggests that anistropic Pauli channel approximations to realistic noise models are pessimistic.  The CMC w approximations also result in lower pseudothresholds. This is in contrast to the isotropic Pauli channel approximation (DC) that yields optimistic pseudothresholds.

The CMCs give more accurate pseudo-threshold estimates than the PCs, as can be seen by comparing their RMS values.  Although we might expect the ``a'' channels to result in better approximations than the ``w'' channels, in general this is not the case.  The most important variation is between the CMCs and the PCs.  In general, however, the ``w'' channels result in lower pseudo-thresholds than the ``a'' channels, which implies that honest approximations at the physical level generate conservative estimates of the threshold.   Finally, we notice that the circuit-based pseudothresholds are quite comparable yielding  pseudothresholds within a factor of two for all of the error models. The DC model representing isotropic depolarizing noise yields the least accurate results and the highest thresholds.

In the previous section, we noticed that the PC a, one of the simplest approximations at the physical level, and one that is not even honest, results in very accurate and practically honest approximations at the corrected logical levels.  In the context of our level-1 pseudo-threshold estimation, this result suggests that we can take a different strategy.  Instead of using the approximate channel at both the physical and logical level to calculate the pseudo-threshold, we can use the target channel at the physical level and the PC a approximation at the logical level.  More generally, we can simulate the realistic noise model in an exact way whenever it is feasible, and in the encoded cases just use the PC a.  If we take this approach, we obtain more accurate state by state pseudo-thresholds for the circuit-based case as seen in Tables \ref{table:summarySteaneADC} and \ref{table:summarySteanePolXY}.

\section{Conclusions} \label{sec:conclusions}

We have studied the feasibility of using approximate error channels at the physical level to simulate the performance of QEC protocols under the influence of non-stabilizer errors.  We have selected the Steane [[7,1,3]] code as a model QEC protocol and have calculated the honesty and accuracy of the Pauli and expanded approximations to realistic non-stabilizer errors. We have also computed the code's pseudothreshold under the different error models.    

Similarly to results recently obtained for distance-3 surface codes \cite{Yu_and_Krysta}, the PC approximations result in lower pseudothreshold values than the realistic error channels.  In contrast, the isotropic DC approximation yields higher pseudothresholds than the target channels in the circuit-based model. Since most thresholds in the literature use two-qubit and one-qubit depolarizing channels, we expect that realistic error models with equivalent fidelity will have slightly lower pseudothresholds in practice.

We have also found that physically honest approximations compose well: they result in honest approximations at the logical level.  Perhaps more interestingly, for both realistic noise models analyzed, the dishonesty of the PC a gets greatly reduced at the corrected logical levels and its accuracy becomes extremely high.  This suggests that, if our goal is to estimate thresholds, then the best protocol is to model the error as realistically as possible at the physical level and use the PC a at the logical level.  

As explained by Puzzuoli \textit{et al.} \cite{Cory2014}, single qubit errors can be separated into errors that deform the Bloch sphere and errors that preserve the Bloch sphere.  Both target error models analyzed in our work deform the Bloch sphere where the PC a yields effectively honest approximations at the logical level for small errors.  Puzzuoli \textit{et al.} found that when the errors are unitary, the PC a generally results in a poor approximation.  In future work, we will compute the level-1 pseudothreshold for the Steane and other codes under unitary errors, to determine if Pauli and expanded approximations still result in pessimistic pseudothreshold estimates.     

\begin{acknowledgments}

This work was supported by the Office of the Director of National Intelligence - Intelligence Advanced Research Projects Activity through ARO contract W911NF-10-1-0231. .All statements of fact, opinion, or conclusions contained herein are those of the authors and should not be construed as representing the official views or policies of IARPA, the ODNI, or the US Government.

\end{acknowledgments}

\bibliographystyle{apsrev}

\begin{thebibliography}{27}
\expandafter\ifx\csname natexlab\endcsname\relax\def\natexlab#1{#1}\fi
\expandafter\ifx\csname bibnamefont\endcsname\relax
  \def\bibnamefont#1{#1}\fi
\expandafter\ifx\csname bibfnamefont\endcsname\relax
  \def\bibfnamefont#1{#1}\fi
\expandafter\ifx\csname citenamefont\endcsname\relax
  \def\citenamefont#1{#1}\fi
\expandafter\ifx\csname url\endcsname\relax
  \def\url#1{\texttt{#1}}\fi
\expandafter\ifx\csname urlprefix\endcsname\relax\def\urlprefix{URL }\fi
\providecommand{\bibinfo}[2]{#2}
\providecommand{\eprint}[2][]{\url{#2}}

\bibitem[{\citenamefont{Poulin}(2006)}]{Poulin_soft_decoding}
\bibinfo{author}{\bibfnamefont{D.}~\bibnamefont{Poulin}},
  \bibinfo{journal}{Phys. Rev. A} \textbf{\bibinfo{volume}{74}},
  \bibinfo{pages}{052333} (\bibinfo{year}{2006}).

\bibitem[{\citenamefont{Goto and Uchikawa}(2013)}]{Goto_soft_decoding}
\bibinfo{author}{\bibfnamefont{H.}~\bibnamefont{Goto}} \bibnamefont{and}
  \bibinfo{author}{\bibfnamefont{H.}~\bibnamefont{Uchikawa}},
  \bibinfo{journal}{Sci. Rep.} \textbf{\bibinfo{volume}{3}},
  \bibinfo{pages}{2044} (\bibinfo{year}{2013}).

\bibitem[{\citenamefont{Suchara et~al.}(2014)\citenamefont{Suchara, Cross, and
  Gambetta}}]{Suchara_leakage}
\bibinfo{author}{\bibfnamefont{M.}~\bibnamefont{Suchara}},
  \bibinfo{author}{\bibfnamefont{A.~W.} \bibnamefont{Cross}}, \bibnamefont{and}
  \bibinfo{author}{\bibfnamefont{J.~M.} \bibnamefont{Gambetta}},
  \bibinfo{journal}{arXiv:1410.8562 [quant-ph]}  (\bibinfo{year}{2014}).

\bibitem[{\citenamefont{Novais and Mucciolo}(2013)}]{Novais_correlated_errors}
\bibinfo{author}{\bibfnamefont{E.}~\bibnamefont{Novais}} \bibnamefont{and}
  \bibinfo{author}{\bibfnamefont{E.~R.}~\bibnamefont{Mucciolo}},
  \bibinfo{journal}{Phys. Rev. Lett.} \textbf{\bibinfo{volume}{110}},
  \bibinfo{pages}{010502} (\bibinfo{year}{2013}).

\bibitem[{\citenamefont{Fowler}(2013)}]{Fowler_leakage}
\bibinfo{author}{\bibfnamefont{A.~G.}~\bibnamefont{Fowler}},
  \bibinfo{journal}{Phys. Rev. A} \textbf{\bibinfo{volume}{88}},
  \bibinfo{pages}{042308} (\bibinfo{year}{2013}).

\bibitem[{\citenamefont{Tomita and Svore}(2014)}]{Yu_and_Krysta}
\bibinfo{author}{\bibfnamefont{Y.}~\bibnamefont{Tomita}} \bibnamefont{and}
  \bibinfo{author}{\bibfnamefont{K.~M.} \bibnamefont{Svore}},
  \bibinfo{journal}{Phys. Rev. A} \textbf{\bibinfo{volume}{90}},
  \bibinfo{pages}{062320} (\bibinfo{year}{2014}).

\bibitem[{\citenamefont{Jouzdani et~al.}(2014)\citenamefont{Jouzdani, Novais,
  Tupitsyn, and Mucciolo}}]{Novais_beyond_single_errors}
\bibinfo{author}{\bibfnamefont{P.}~\bibnamefont{Jouzdani}},
  \bibinfo{author}{\bibfnamefont{E.}~\bibnamefont{Novais}},
  \bibinfo{author}{\bibfnamefont{I.~S.} \bibnamefont{Tupitsyn}},
  \bibnamefont{and} \bibinfo{author}{\bibfnamefont{E.~R.}
  \bibnamefont{Mucciolo}}, \bibinfo{journal}{Phys. Rev. A}
  \textbf{\bibinfo{volume}{90}}, \bibinfo{pages}{042315}
  (\bibinfo{year}{2014}).

\bibitem[{\citenamefont{Gottesman}(1997)}]{Gottesman_thesis}
\bibinfo{author}{\bibfnamefont{D.}~\bibnamefont{Gottesman}}, Ph.D. thesis,
  \bibinfo{school}{California Institute of Technology} (\bibinfo{year}{1997}).

\bibitem[{\citenamefont{Aaronson and Gottesman}(2004)}]{Aaronson}
\bibinfo{author}{\bibfnamefont{S.}~\bibnamefont{Aaronson}} \bibnamefont{and}
  \bibinfo{author}{\bibfnamefont{D.}~\bibnamefont{Gottesman}},
  \bibinfo{journal}{Phys. Rev. A} \textbf{\bibinfo{volume}{70}},
  \bibinfo{pages}{052328} (\bibinfo{year}{2004}).

\bibitem[{\citenamefont{Magesan et~al.}(2013)\citenamefont{Magesan, Puzzuoli,
  Granade, and Cory}}]{Cory}
\bibinfo{author}{\bibfnamefont{E.}~\bibnamefont{Magesan}},
  \bibinfo{author}{\bibfnamefont{D.}~\bibnamefont{Puzzuoli}},
  \bibinfo{author}{\bibfnamefont{C.~E.} \bibnamefont{Granade}},
  \bibnamefont{and} \bibinfo{author}{\bibfnamefont{D.~G.} \bibnamefont{Cory}},
  \bibinfo{journal}{Phys. Rev. A} \textbf{\bibinfo{volume}{87}},
  \bibinfo{pages}{012324} (\bibinfo{year}{2013}).

\bibitem[{\citenamefont{Guti\'errez et~al.}(2013)\citenamefont{Guti\'errez,
  Svec, Vargo, and Brown}}]{PRA_us}
\bibinfo{author}{\bibfnamefont{M.}~\bibnamefont{Guti\'errez}},
  \bibinfo{author}{\bibfnamefont{L.}~\bibnamefont{Svec}},
  \bibinfo{author}{\bibfnamefont{A.}~\bibnamefont{Vargo}}, \bibnamefont{and}
  \bibinfo{author}{\bibfnamefont{K.~R.} \bibnamefont{Brown}},
  \bibinfo{journal}{Phys. Rev. A} \textbf{\bibinfo{volume}{87}},
  \bibinfo{pages}{030302} (\bibinfo{year}{2013}).

\bibitem[{\citenamefont{Steane}(1996)}]{Steane1996}
\bibinfo{author}{\bibfnamefont{A.~M.} \bibnamefont{Steane}},
  \bibinfo{journal}{Phys. Rev. Lett.} \textbf{\bibinfo{volume}{77}},
  \bibinfo{pages}{793} (\bibinfo{year}{1996}).

\bibitem[{\citenamefont{Steane}(2003)}]{SteanePRA2003}
\bibinfo{author}{\bibfnamefont{A.~M.} \bibnamefont{Steane}},
  \bibinfo{journal}{Phys. Rev. A} \textbf{\bibinfo{volume}{68}},
  \bibinfo{pages}{042322} (\bibinfo{year}{2003}).

\bibitem[{\citenamefont{Svore et~al.}(2007)\citenamefont{Svore, DiVincenzo, and
  Terhal}}]{SvoreQIC2007}
\bibinfo{author}{\bibfnamefont{K.~M.} \bibnamefont{Svore}},
  \bibinfo{author}{\bibfnamefont{D.~P.} \bibnamefont{DiVincenzo}},
  \bibnamefont{and} \bibinfo{author}{\bibfnamefont{B.~M.}
  \bibnamefont{Terhal}}, \bibinfo{journal}{Quantum Information {\&}
  Computation} \textbf{\bibinfo{volume}{7}}, \bibinfo{pages}{297}
  (\bibinfo{year}{2007}).

\bibitem[{\citenamefont{Metodi et~al.}({2005})\citenamefont{Metodi, Thaker,
  Cross, Chong, and Chuang}}]{MetodiMicro2005}
\bibinfo{author}{\bibfnamefont{T.~S.}~\bibnamefont{Metodi}},
  \bibinfo{author}{\bibfnamefont{D.~D.}~\bibnamefont{Thaker}},
  \bibinfo{author}{\bibfnamefont{A.~W.}~\bibnamefont{Cross}},
  \bibinfo{author}{\bibfnamefont{F.~T.}~\bibnamefont{Chong}}, \bibnamefont{and}
  \bibinfo{author}{\bibfnamefont{I.~L.}~\bibnamefont{Chuang}}, in
  \emph{\bibinfo{booktitle}{MICRO-38: Proc. 38TH Annual IEEE/ACM Int. Symp. on
  Microarchitecture}} (\bibinfo{year}{{2005}}), pp.
  \bibinfo{pages}{{305--316}}.

\bibitem[{\citenamefont{Tomita et~al.}(2013)\citenamefont{Tomita, Guti\'errez,
  Kabytayev, Brown, Hutsel, Morris, Stevens, and Mohler}}]{TomitaPRA2013}
\bibinfo{author}{\bibfnamefont{Y.}~\bibnamefont{Tomita}},
  \bibinfo{author}{\bibfnamefont{M.}~\bibnamefont{Guti\'errez}},
  \bibinfo{author}{\bibfnamefont{C.}~\bibnamefont{Kabytayev}},
  \bibinfo{author}{\bibfnamefont{K.~R.}~\bibnamefont{Brown}},
  \bibinfo{author}{\bibfnamefont{M.~R.}~\bibnamefont{Hutsel}},
  \bibinfo{author}{\bibfnamefont{A.~P.}~\bibnamefont{Morris}},
  \bibinfo{author}{\bibfnamefont{K. E.}~\bibnamefont{Stevens}}, \bibnamefont{and}
  \bibinfo{author}{\bibfnamefont{G.}~\bibnamefont{Mohler}},
  \bibinfo{journal}{Phys. Rev. A} \textbf{\bibinfo{volume}{88}},
  \bibinfo{pages}{042336} (\bibinfo{year}{2013}).


\bibitem[{\citenamefont{Weinstein}(2014)}]{WeinsteinPRA2014}
\bibinfo{author}{\bibfnamefont{Y.~S.} \bibnamefont{Weinstein}},
  \bibinfo{journal}{Phys. Rev. A} \textbf{\bibinfo{volume}{89}},
  \bibinfo{pages}{020301} (\bibinfo{year}{2014}).

\bibitem[{\citenamefont{Abu-Nada, Fortescue, Byrd}(2014)}]{AbuNadaPRA2014}
\bibinfo{author}{\bibfnamefont{A.}~\bibnamefont{Abu-Nada}},
  \bibinfo{author}{\bibfnamefont{B.}~\bibnamefont{Fortescue}},
  \bibnamefont{and}
  \bibinfo{author}{\bibfnamefont{M.}~\bibnamefont{Byrd}},
  \bibinfo{journal}{Phys. Rev. A} \textbf{\bibinfo{volume}{89}},
  \bibinfo{pages}{062304} (\bibinfo{year}{2014}).


\bibitem[{\citenamefont{Anderson et~al.}(2014)\citenamefont{Anderson,
  Duclos-Cianci, and Poulin}}]{AndersonPRL2014}
\bibinfo{author}{\bibfnamefont{J.~T.} \bibnamefont{Anderson}},
  \bibinfo{author}{\bibfnamefont{G.}~\bibnamefont{Duclos-Cianci}},
  \bibnamefont{and} \bibinfo{author}{\bibfnamefont{D.}~\bibnamefont{Poulin}},
  \bibinfo{journal}{Phys. Rev. Lett.} \textbf{\bibinfo{volume}{113}},
  \bibinfo{pages}{080501} (\bibinfo{year}{2014}).

\bibitem[{\citenamefont{Nigg et~al.}(2014)\citenamefont{Nigg, M\"{u}ller,
  Martinez, Schindler, Hennrich, Monz, Martin-Delgado, and
  Blatt}}]{NiggScience2014}
\bibinfo{author}{\bibfnamefont{D.}~\bibnamefont{Nigg}},
  \bibinfo{author}{\bibfnamefont{M.}~\bibnamefont{M\"{u}ller}},
  \bibinfo{author}{\bibfnamefont{E.~A.} \bibnamefont{Martinez}},
  \bibinfo{author}{\bibfnamefont{P.}~\bibnamefont{Schindler}},
  \bibinfo{author}{\bibfnamefont{M.}~\bibnamefont{Hennrich}},
  \bibinfo{author}{\bibfnamefont{T.}~\bibnamefont{Monz}},
  \bibinfo{author}{\bibfnamefont{M.~A.} \bibnamefont{Martin-Delgado}},
  \bibnamefont{and} \bibinfo{author}{\bibfnamefont{R.}~\bibnamefont{Blatt}},
  \bibinfo{journal}{Science} \textbf{\bibinfo{volume}{345}},
  \bibinfo{pages}{302} (\bibinfo{year}{2014}).

\bibitem[{\citenamefont{Puzzuoli et~al.}(2014)\citenamefont{Puzzuoli, Granade,
  Haas, Criger, Magesan, and Cory}}]{Cory2014}
\bibinfo{author}{\bibfnamefont{D.}~\bibnamefont{Puzzuoli}},
  \bibinfo{author}{\bibfnamefont{C.}~\bibnamefont{Granade}},
  \bibinfo{author}{\bibfnamefont{H.}~\bibnamefont{Haas}},
  \bibinfo{author}{\bibfnamefont{B.}~\bibnamefont{Criger}},
  \bibinfo{author}{\bibfnamefont{E.}~\bibnamefont{Magesan}}, \bibnamefont{and}
  \bibinfo{author}{\bibfnamefont{D.~G.} \bibnamefont{Cory}},
  \bibinfo{journal}{Phys. Rev. A} \textbf{\bibinfo{volume}{89}},
  \bibinfo{pages}{022306} (\bibinfo{year}{2014}).

\bibitem[{\citenamefont{Geller and Zhou}(2013)}]{Geller_and_Zhou}
\bibinfo{author}{\bibfnamefont{M.~R.} \bibnamefont{Geller}} \bibnamefont{and}
  \bibinfo{author}{\bibfnamefont{Z.}~\bibnamefont{Zhou}},
  \bibinfo{journal}{Phys. Rev. A} \textbf{\bibinfo{volume}{88}},
  \bibinfo{pages}{012314} (\bibinfo{year}{2013}).

\bibitem[{\citenamefont{van Dam and Howard}(2009)}]{vanDam2009}
\bibinfo{author}{\bibfnamefont{W.}~\bibnamefont{van Dam}} \bibnamefont{and}
  \bibinfo{author}{\bibfnamefont{M.}~\bibnamefont{Howard}},
  \bibinfo{journal}{Phys. Rev. Lett.} \textbf{\bibinfo{volume}{103}},
  \bibinfo{pages}{170504} (\bibinfo{year}{2009}).

\bibitem[{\citenamefont{Grace et~al.}(2010)\citenamefont{Grace, Dominy, Kosut,
  Brif, and Rabitz}}]{Distance_meas}
\bibinfo{author}{\bibfnamefont{M.~D.} \bibnamefont{Grace}},
  \bibinfo{author}{\bibfnamefont{J.}~\bibnamefont{Dominy}},
  \bibinfo{author}{\bibfnamefont{R.~L.} \bibnamefont{Kosut}},
  \bibinfo{author}{\bibfnamefont{C.}~\bibnamefont{Brif}}, \bibnamefont{and}
  \bibinfo{author}{\bibfnamefont{H.}~\bibnamefont{Rabitz}},
  \bibinfo{journal}{New Journal of Physics} \textbf{\bibinfo{volume}{12}},
  \bibinfo{pages}{015001} (\bibinfo{year}{2010}).

\bibitem[{\citenamefont{Gilchrist et~al.}(2005)\citenamefont{Gilchrist,
  Langford, and Nielsen}}]{Chuang_distance}
\bibinfo{author}{\bibfnamefont{A.}~\bibnamefont{Gilchrist}},
  \bibinfo{author}{\bibfnamefont{N.~K.} \bibnamefont{Langford}},
  \bibnamefont{and} \bibinfo{author}{\bibfnamefont{M.~A.}
  \bibnamefont{Nielsen}}, \bibinfo{journal}{Phys. Rev. A}
  \textbf{\bibinfo{volume}{71}}, \bibinfo{pages}{062310}
  (\bibinfo{year}{2005}).

\bibitem[{\citenamefont{Chuang and Nielsen}(1997)}]{Pauli_twirling_Chuang}
\bibinfo{author}{\bibfnamefont{I.~L.} \bibnamefont{Chuang}} \bibnamefont{and}
  \bibinfo{author}{\bibfnamefont{M.~A.} \bibnamefont{Nielsen}},
  \bibinfo{journal}{J. Mod. Opt.} \textbf{\bibinfo{volume}{44}},
  \bibinfo{pages}{2455} (\bibinfo{year}{1997}).

\bibitem[{\citenamefont{Divincenzo and Shor}(1996)}]{DivincenzoPRL1996}
\bibinfo{author}{\bibfnamefont{D.~P.}~\bibnamefont{DiVincenzo}}
\bibnamefont{and}
  \bibinfo{author}{\bibfnamefont{P.~W.} \bibnamefont{Shor}},
  \bibinfo{journal}{Phys. Rev. Lett.} \textbf{\bibinfo{volume}{77}},
  \bibinfo{pages}{3260} (\bibinfo{year}{1996}).

\bibitem[{\citenamefont{Steane}(1998)}]{Steane1997}
\bibinfo{author}{\bibfnamefont{A.}~\bibnamefont{Steane}},
  \bibinfo{journal}{Fortsch. Phys.} \textbf{\bibinfo{volume}{46}},
  \bibinfo{pages}{443} (\bibinfo{year}{1998}).

\end{thebibliography}

\end{document}